\documentclass[notitlepage,amsmath,amssymb, aps,superscriptaddress]{revtex4-1}
\usepackage{graphicx,color}

\begin{document}
\noindent{\large SNOWMASS 2021 White Paper contribution} \\




\title{\it \huge \textbf{Combined signatures of heavy Higgses and vectorlike fermions at the HL-LHC}}

\author{Radovan Dermisek}
\email{dermisek@indiana.edu}
\affiliation{Physics Department, Indiana University, Bloomington, IN 47405, USA}
\author{Junichiro Kawamura}
\email{jkawa@ibs.re.kr}
\affiliation{Center for Theoretical Physics of the Universe, Institute for Basic Science, Daejeon 34126, Korea}
\author{Enrico Lunghi}
\email{elunghi@indiana.edu}
\affiliation{Physics Department, Indiana University, Bloomington, IN 47405, USA}
\author{Navin McGinnis}
\email{nmcginnis@triumf.ca}
\affiliation{TRIUMF, 4004 Westbrook Mall, Vancouver, BC, Canada, V6T 2A3}
\author{Seodong Shin}
\email{sshin@jbnu.ac.kr}
\affiliation{Department of Physics, Jeonbuk National University, Jeonju, Jeonbuk 54896, Korea}

\begin{abstract}
\noindent\textbf{Executive Summary:} In extensions of two Higgs doublet models with vectorlike quarks and leptons the decays of new scalars or fermions can be altered by the presence of the other, depending on the hierarchy of new physics. If new fermions are heavier than Higgses, their decays may be dominated by cascade decays through charged and neutral heavy Higgs bosons. In the opposite hierarchy, the decay of a heavy Higgs can lead to single production of vectorlike quarks and leptons when Yukawa couplings to SM fermions are present. We explore the resulting novel collider signatures in either case and present projected sensitivities of future runs of the LHC to the masses and branching ratios of new fermions and heavy Higgses. Importantly, we discuss kinematic challenges of the signatures and suggest strategic observables which could be adopted in future analyses. Interpreting the prospective reach of the LHC in the context of a type-II 2HDM, we find that in the presence of vectorlike fermions heavy Higgs boson masses slightly above 2 TeV can be explored at the HL-LHC even at low $\tan\beta$. For heavy Higgs decays through vectorlike leptons, this reach extends to slightly above 3 TeV for $\tan\beta \gtrsim 10$. Further, these search strategies imply a reach for vectorlike quark and lepton masses close to 2.4 TeV and 1.5 TeV, respectively. This article is partly based on Refs.~\cite{Dermisek:2015hue,Dermisek:2016via,Dermisek:2019vkc,Dermisek:2019heo,Dermisek:2020gbr,Dermisek:2021zjd}.
\end{abstract}

\maketitle

\section{Introduction}

Among the simplest extensions of the Standard Model (SM) are models with an extended Higgs sector or extra generations of fermions. Well-motivated UV completions include supersymmetry, two Higgs doublet models, and composite Higgs models.
Unlike the SM fermions, new fermions are expected to be vectorlike in perturbative scenarios due to the strong experimental constraints on the Higgs measurements at the Large Hadron Collider (LHC) and electroweak precision measurements at the Large Electron-Positron Collider~\cite{BarShalom:2012ms}.

Conventional searches for BSM physics typically focus on the production of individual species of particles followed by their decays to SM bosons and fermions. For instance, searches for heavy Higgs bosons are expected to probed masses $\lesssim 1.8$ TeV at the High Luminosity LHC (HL-LHC) with the full data, 3 ab$^{-1}$, whereas searches for vectorlike quarks are expected to reach $\lesssim 1.5$ TeV~\cite{CidVidal:2018eel}.
Searches for vectorlike leptons from their Drell-Yan productions are not expected to probe masses much above  $1$ TeV due to the small cross sections~\cite{Dermisek:2014qca}.
Compared to the above separate investigations, we point out that searches for combined signatures of both BSM Higgs bosons and vectorlike fermions (including leptons) can increase the experimental sensitivities, based on the recent analysis results in Refs.~\cite{Dermisek:2014cia,Dermisek:2015hue,Dermisek:2015oja,Dermisek:2015vra,Dermisek:2015vra,Dermisek:2016via,Dermisek:2019heo,Dermisek:2019vkc,Dermisek:2020gbr}.

In the following sections, we present a summary of results providing projections for the reach of combined signatures for heavy Higgs bosons and vectorlike fermions at the HE/HL-LHC. In section~\ref{sec:VLQ}, we summarize recent analyses which study the decay pattern of vectorlike quarks and charged and neutral heavy Higgs bosons in both limits when heavy Higgses are heavier and lighter than new quarks. In section~\ref{sec:VLL}, we summarize similar studies with vectorlike leptons.

\section{vectorlike quark signatures}
\label{sec:VLQ}

We focus on a two Higgs doublet model extended with $SU(2)$ doublet- and singlet- vectorlike quarks and leptons. We assume that the new fermion multiplets have analogous quantum numbers to their doublet and singlet SM counterparts. See Table~\ref{table:fieldcontents} for a summary of the field content and charge conventions we consider. An additional $Z_{2}$ symmetry is assumed in order to enforce that vectorlike quarks and leptons couple to the two Higgs doublets as in the type-II model i.e. down-type quarks and charged lepton singlets couple to $H_{d}$ and up-type quarks and neutral singlet leptons couple to $H_{u}$. For further details of the model we consider see~\cite{Dermisek:2019vkc,Dermisek:2021ajd}. Distinct experimental signatures for heavy fermions, heavy Higgses, or both in these scenarios can be distinguished depending on the hierarchy between new fermions and Higgses. In this section, we discuss combined signatures of vectorlike quarks and heavy Higgses both in the case that new quarks are heavier than the Higgs sector and vice versa. 
\subsection{Vectorlike quark cascade decays}
\begin{table}[t]
\begin{center}
\resizebox{\columnwidth}{!}{
\begin{tabular}{c c c c c c c c c c c c c c}
\hline
\hline
 & ~~$\ell^i_L$ & ~~$e^i_R$ & ~~$q^i_L$ & ~~$u^i_R$ & ~~$d^i_R$ & ~~$L_{L,R}$ & ~~$E_{L,R}$ & ~~$N_{L,R}$ & ~~$Q_{L,R}$ & ~~$T_{L,R}$ & ~~$B_{L,R}$ & ~~$H_d$ & ~~ $H_u$\\
\hline
SU(2)$_{\rm L}$ & ~~\bf 2 & ~~\bf 1 & ~~\bf 2 & ~~\bf 1 & ~~\bf 1 & ~~\bf 2 & ~~\bf 1 & ~~\bf 1 & ~~\bf 2 & ~~\bf 1 & ~~\bf 1 & ~~\bf 2 & ~~\bf 2 \\
U(1)$_{\rm Y}$ & ~~$-\frac12$ & ~~-1 & ~~$\frac16$ & ~~$\frac23$ & ~~-$\frac13$ & ~~$-\frac12$ & ~~-1 & ~~0  & ~~$\frac16$ & ~~$\frac23$ & ~~-$\frac13$ & ~~$\frac12$ & ~~-$\frac12$ \\
Z$_2$ & ~~+ & ~~- & ~~+ & ~~+ & ~~-- & ~~+ & ~~- & ~~+ & ~~+ & ~~+ & ~~-- & ~~-- & ~~+ \\
\hline
\hline
\end{tabular}}
\end{center}
    \caption{Quantum numbers under $SU(2)_{L}\times U(1)_{Y}$ of standard model leptons and quarks ($\ell^i_L$, $e^i_R$, $q^i_L, u^i_R, d^i_R$ for $i=1,2,3$), extra vectorlike leptons and quarks and the two Higgs doublets. $Z_{2}$ charges assigned to couple fermions to the Higgs doublets as in the type-II model are given in the last row.
    }
\label{table:fieldcontents}
\end{table}

\begin{figure}[t]
\centering
\includegraphics[width=0.45\textwidth]{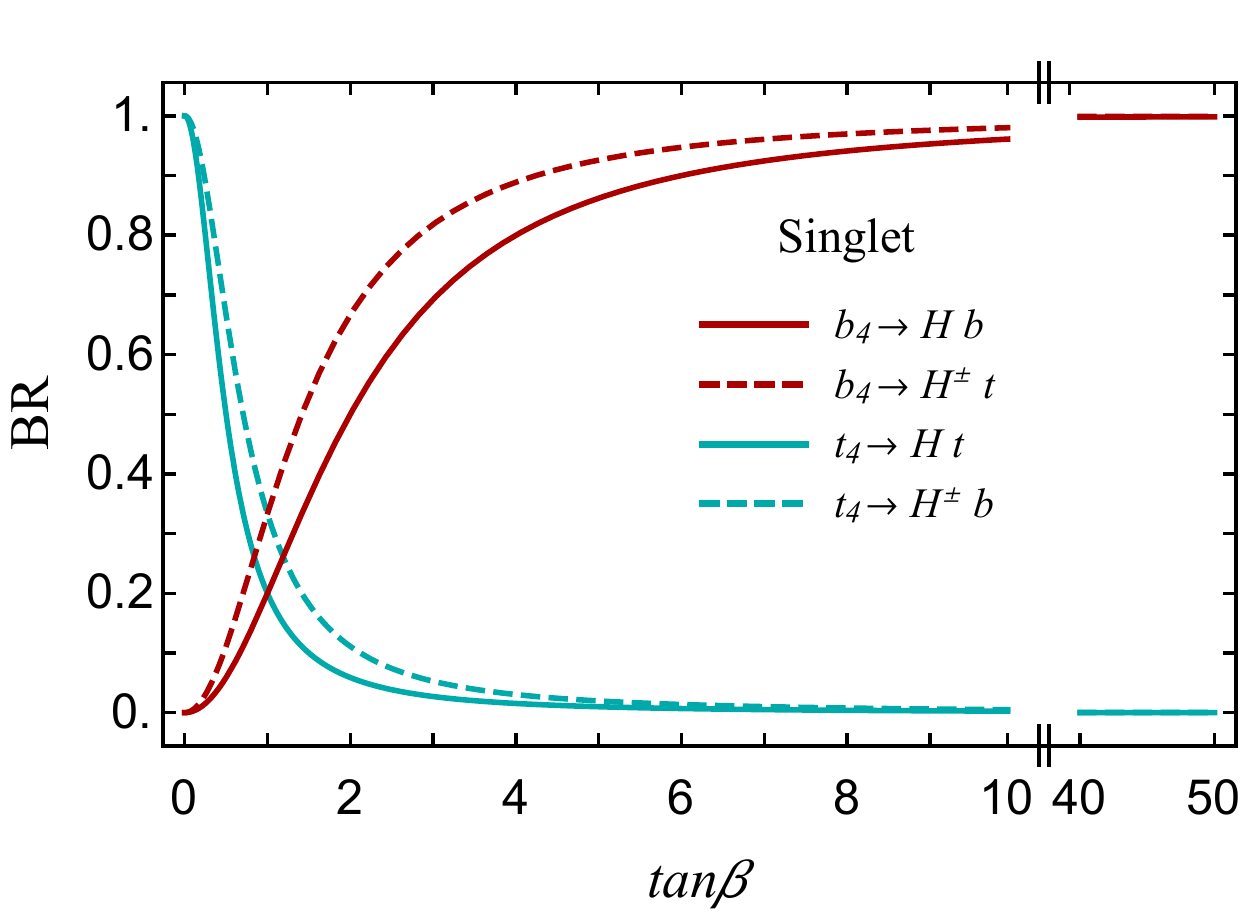}
\includegraphics[width=0.45\textwidth]{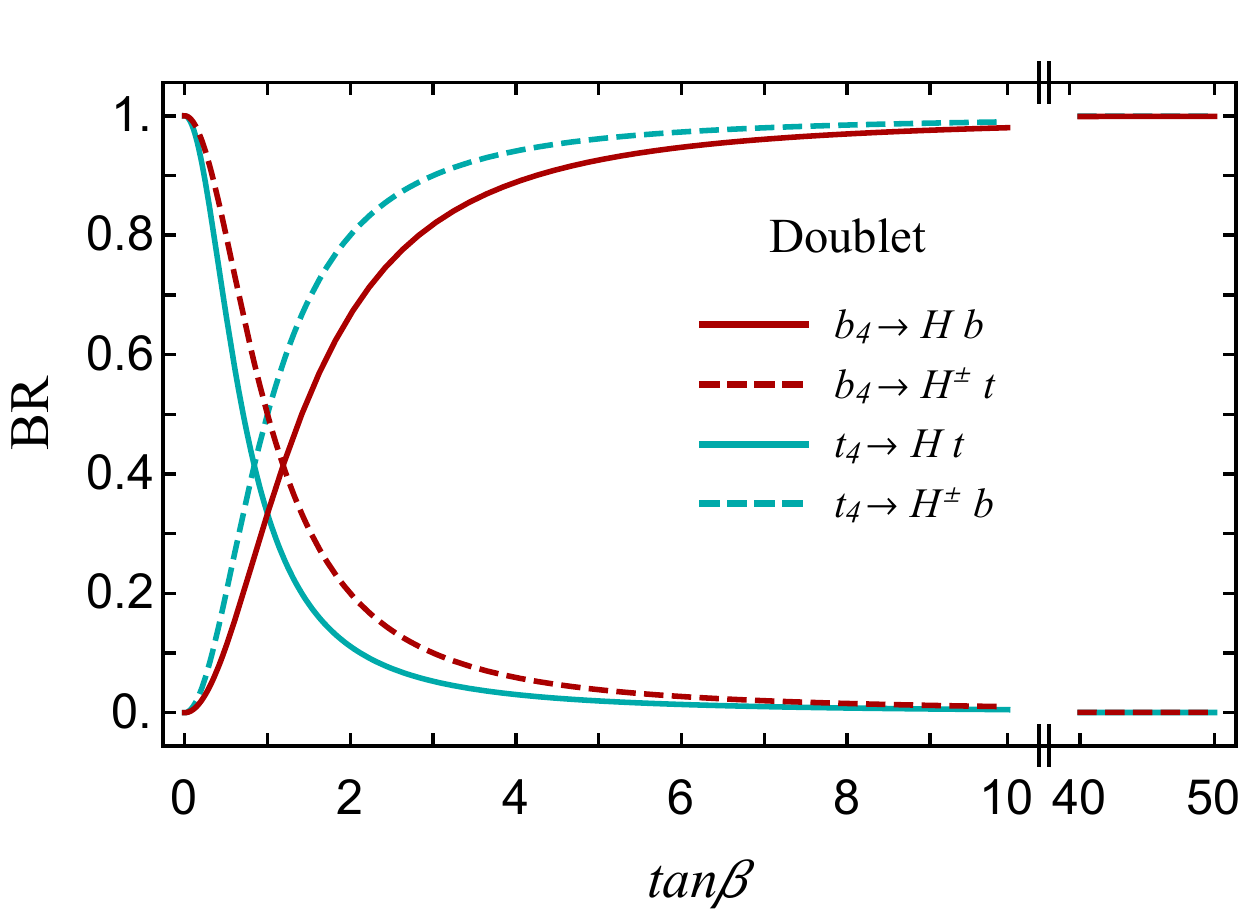}
\caption{Leading order expectation of singlet-like (left) and doublet-like (right) $b_4$ and $t_4$ branching ratios to charged and neutral heavy Higgses as functions of $\tan\beta$. Branching ratios involving the CP-odd heavy Higgs, $A$, are identical to those shown with $H$. Each branching ratio is obtained assuming only one heavy Higgs decay channel is kinematically open. This plot is taken from~\cite{Dermisek:2021zjd}}
\label{fig:singlet_doublet_BR}
\end{figure}

Mixing between vectorlike and SM fermions via couplings to the two Higgs doublets introduces several new decay modes of top- ($t_{4}$) and bottom-like ($b_{4}$) vectorlike quarks. Assuming that vectorlike fermions are heavier than the Higgs spectrum, we show in Fig.~\ref{fig:singlet_doublet_BR} the typical branching ratios of top- and bottom-like vectorlike quarks to heavy charged and neutral Higgs bosons as a function of the ratio of vacuum expectation values of the two Higgs doublets, $\tan\beta$. In the left (right) panel, we show the $\tan\beta$ dependence when the $t_{4}$ and $b_{4}$ mass eigenstates are mostly singlet-(doublet) like. In either case, there exists regions of parameters where decays of vectorlike quarks to heavy Higgses are expected to dominate over the decays to the $W$, $Z$, or SM $h$ bosons. Importantly, this feature persists regardless of the size of the coupling which mixes heavy quarks and SM quarks. After EWSB the same coupling controls the flavor-violating decays of heavy quarks to $W$, $Z$, or SM $h$ bosons and SM quarks. Thus, the branching ratios of heavy quarks are independent of these couplings at leading order and the only free parameter left is $\tan\beta$.

\begin{figure}[t]
\centering
\includegraphics[width=0.6\linewidth]{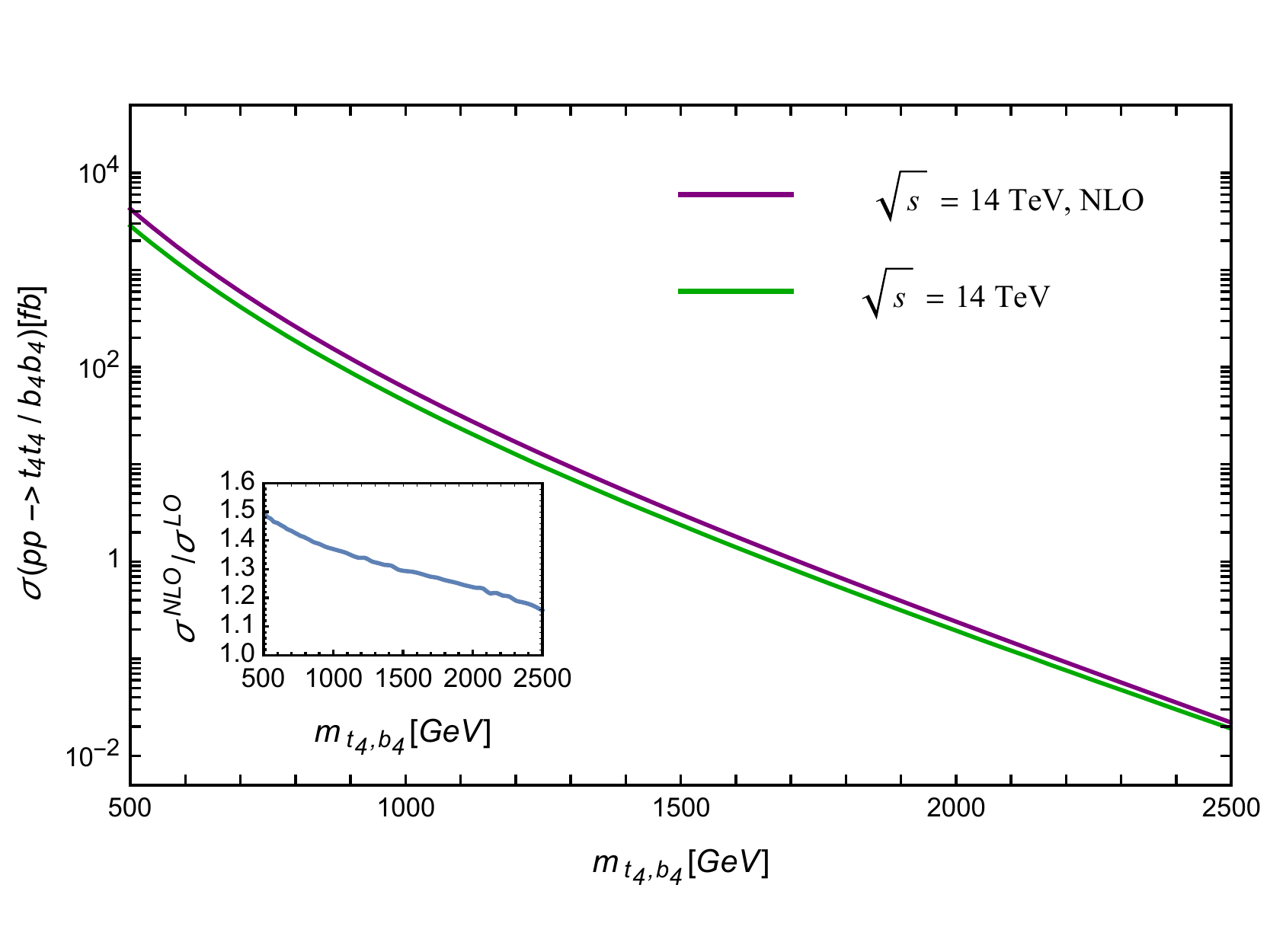}
\caption{Pair production cross section for top- or bottom-like vector-like quarks at the 14 TeV LHC with respect to the quark masses. We show the LO result obtained with {\tt MadGraph5}~\cite{Alwall:2014hca}(green), the NLO result obtained with {\tt Top++}~\cite{Czakon:2011xx}(purple), and the relative enhancement between the two (inset). This plot is taken from~\cite{Dermisek:2021zjd}.}
\label{fig:xsection}
\end{figure}

The production of heavy Higgses in the decay modes of vectorlike quarks has the additional advantage that heavy Higgses are effectively produced via QCD-sized cross sections in the regions of parameters where the branching ratios to heavy Higgses are dominant. Further, these cross sections are model independent and only depend on the mass of the produced vectorlike quark at the LHC. In Fig.~\ref{fig:xsection}, we show the $t_{4}$ or $b_{4}$ pair production cross section at the LHC for $\sqrt{s}=14$ TeV as a function of the vectorlike quark mass. We show the leading order result (green curve) compared to the cross section at NLO (purple curve), and the relative enhancement between the two (inset).

From the discussion above, two main kinematic signatures standout as promising signals for vectorlike quark cascade decays at the LHC which are determined by the pattern of subsequent heavy Higgs decays. In a two Higgs doublet model type-II the dominant decay modes of heavy charged and neutral Higgs bosons are to SM top and bottom quarks. Thus, at the level of cross sections, signatures with multiple top and bottom quarks in the final state are expected to produce the largest rates. However, as will be discussed in the following section, it is expected that this signature will also suffer from large QCD backgrounds. An alternative to this strategy is to focus on signatures resulting from the subdominant decay mode of heavy Higgses to $\tau$ leptons. We summarize these search strategies and projections at the 14 TeV HL-LHC in the folliowing sections.

\subsubsection{6$b$ signatures}
\begin{figure}[t]
\begin{minipage}{6in}
\centering
	\raisebox{-0.5\height}{ \includegraphics[scale=0.35]{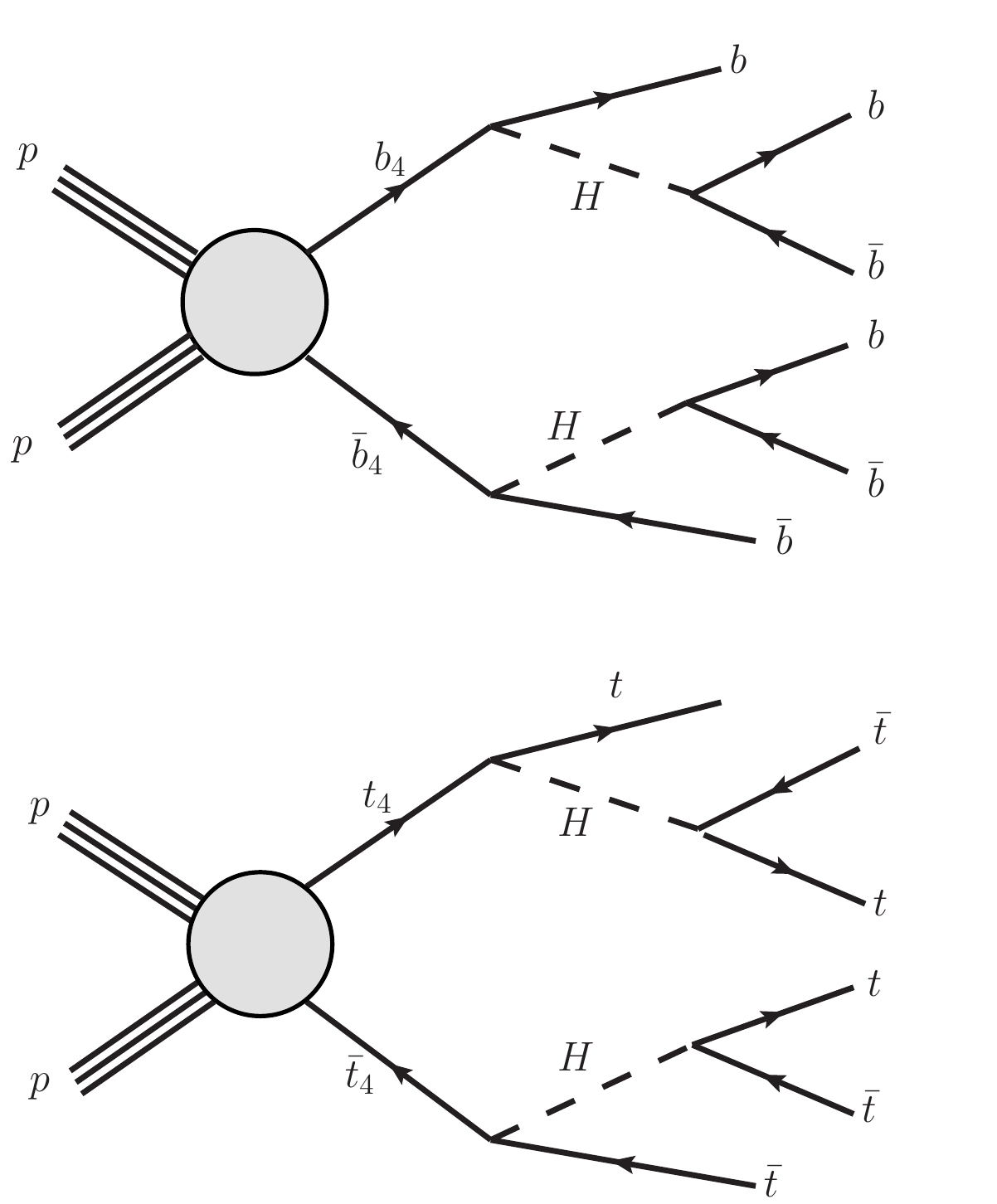} }
  \hspace*{0.1in
  	\raisebox{-0.5\height}{ \includegraphics[scale=0.35]{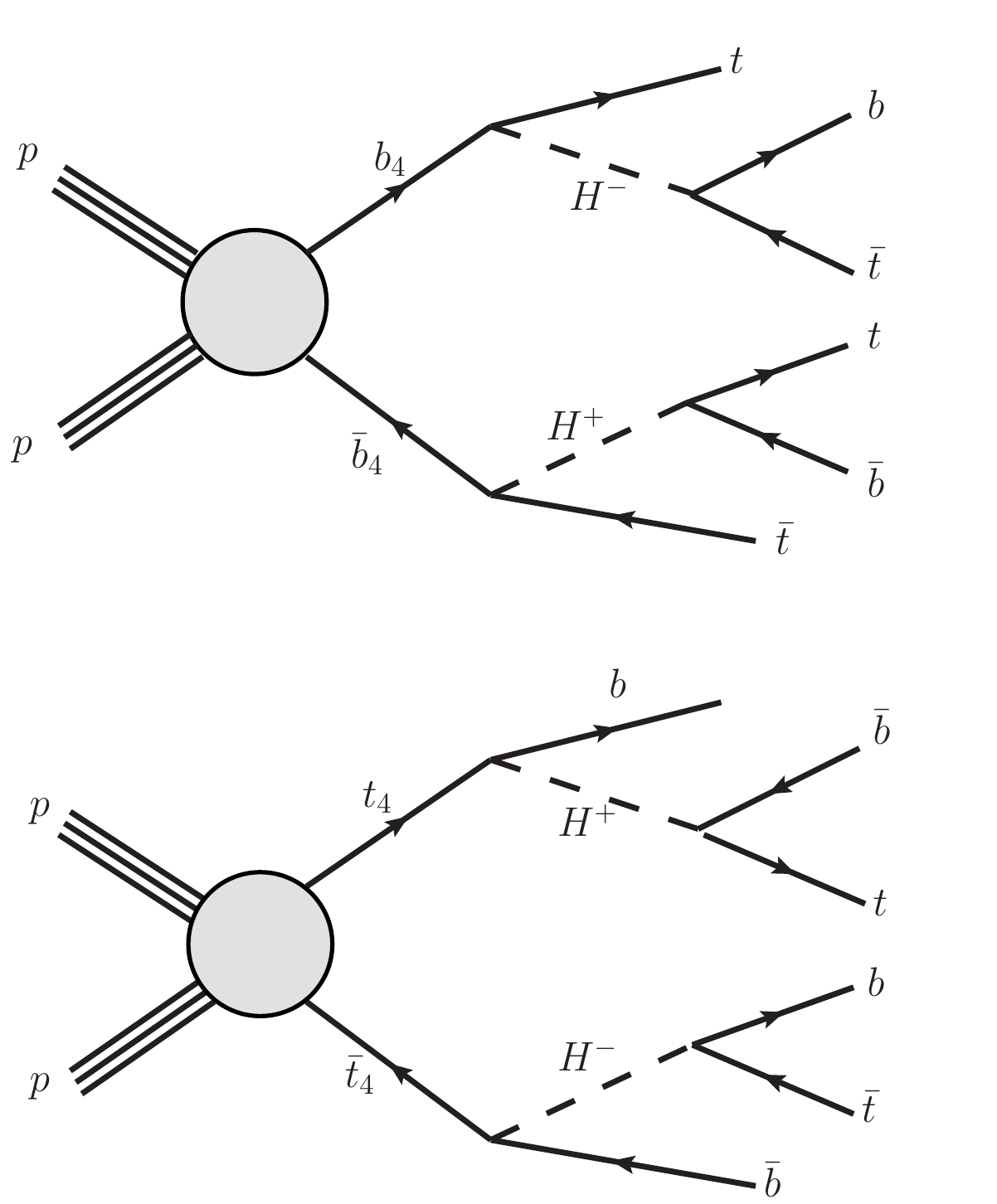} }
    \hspace*{0.1in}
	\raisebox{-0.5\height}{ \includegraphics[scale=0.35]{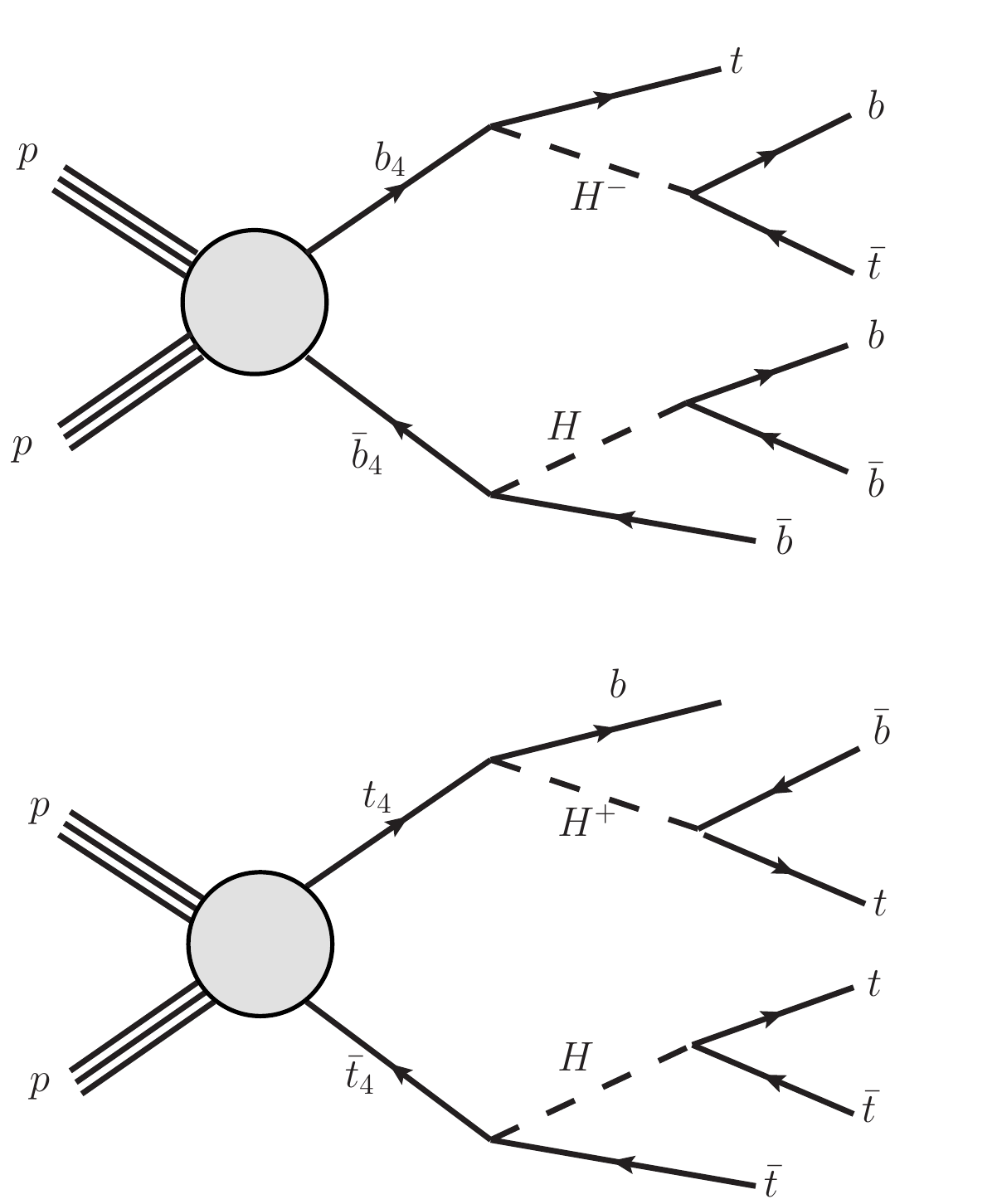}}
\caption{Dominant cascade decays resulting from pair produced vectorlike top and bottom quarks decaying through neutral Higgs bosons (left), charged Higgs bosons (middle), and mixed cases of one decaying through a neutral and the other through a charged Higgs boson (right). This figure is taken from~\cite{Dermisek:2020gbr}.}
\label{fig:diagrams}
}
\end{minipage}
\end{figure}

In Fig.~\ref{fig:diagrams}, we show the possible signatures resulting from cascade decays of vectorlike quarks through heavy Higgs to multiple top and bottom quarks at the LHC. Note that we have disregarded signatures resulting from $t_{4}\to Ht \to b\bar bt$ and $b_{4}\to Hb \to t\bar tt$ decays as the combined branching ratio is subdominant compared to those we show. Despite the large number of possible final states, considering the fact that top quarks mostly decay to bottom quarks, all the resulting signatures can be captured by analyses focusing on tagging multiple $b$-jets. This strategy was used in \cite{Dermisek:2020gbr} and we summarize the results here.

In contrast to searches involving $h\to b\bar b$ or $h\to t\bar t$ decays, the presence of heavy Higgses in the cascade decay chain results in widely separated $b$-jets and large hadronic energies as kinematic handles on the signal. However, irreducible QCD multijet backgrounds are difficult to simulate in this case. The kinematic variable
\begin{equation}
H_{Tb}\equiv\sum_{j\in b} |p_{T}(j)|
\end{equation}
quantifying the total hadronic energy carried by the $b$-jets was introduced in~\cite{Dermisek:2020gbr} as a means of reducing higher order corrections from multijet backgrounds. Another challenge arises from distinguishing $b$-jets produced from gluon splitting~\cite{Goncalves:2015prv}.

\begin{figure}[t]
\centering
\includegraphics[width=0.6\linewidth]{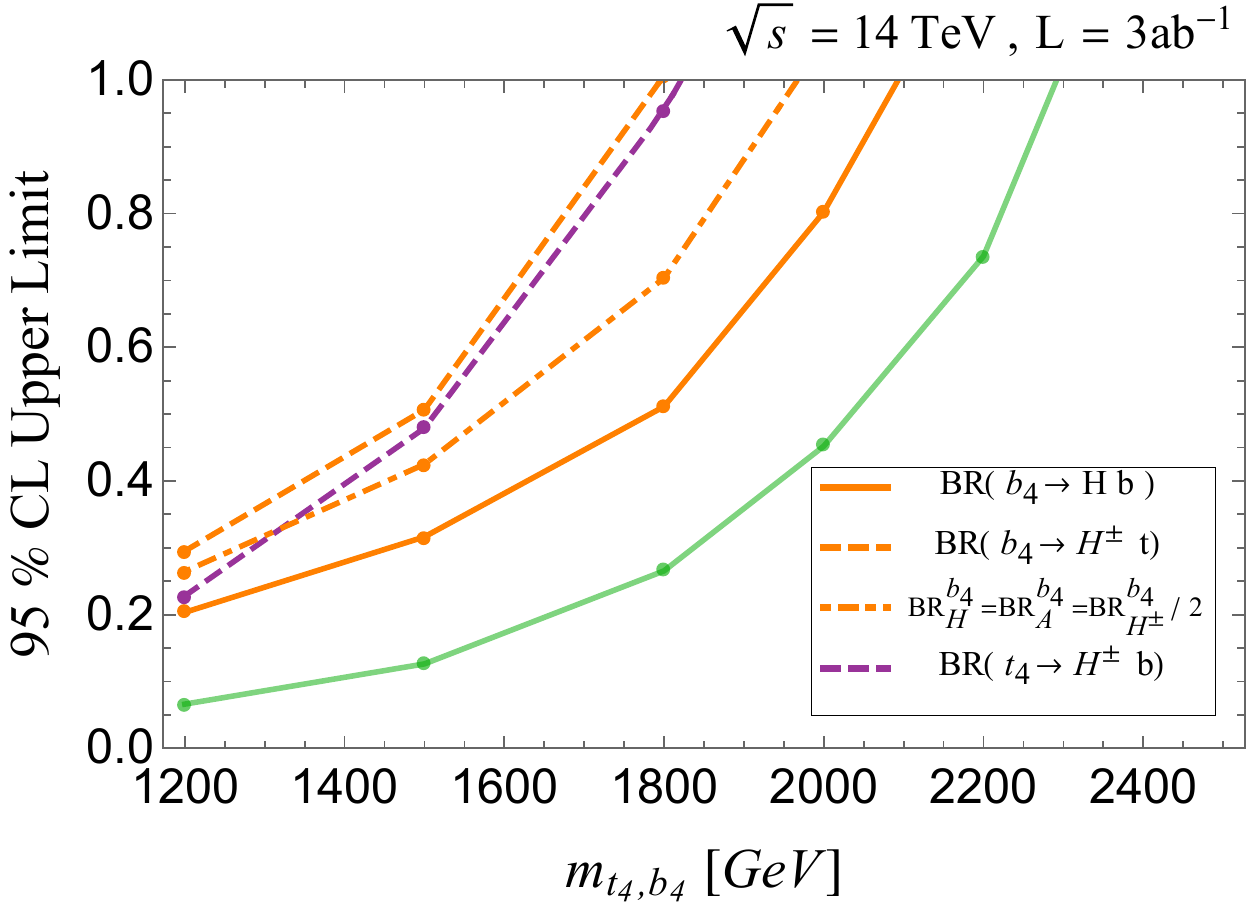}
\caption{95\% CL upper limits on $t_{4}$ and $b_{4}$ cascade decays for the 14 TeV HL-LHC. The projected limits on the ${\rm BR}(b_{4}\to Hb)$ decay using the 5b analysis in ref.~\cite{Dermisek:2020gbr} is shown in green assuming ${\rm BR}(H\to b\bar{b})=0.9$, whereas limits on charged, neutral, and mixed heavy Higgs decays of $t_{4}$ and $b_{4}$ obtained using the $\tau$-jet analysis are shown in dashed orange and purple, solid orange, and dotted orange, respectively. This plot is taken from~\cite{Dermisek:2021zjd}.}
\label{fig:BR_bounds}
\end{figure}

In~\cite{Dermisek:2020gbr}, we explored analyses based on tagging 4 and 5 large $p_{T}$ $b$-jets using a simple cut-and-count strategy with $H_{Tb}$ as the main kinematic variable. We show the optimal reach of this strategy with the green curve in Fig.~\ref{fig:BR_bounds}. For a given vectorlike quark mass we find a 95\% upper limit on the branching ratio of vectorlike quarks to heavy Higgses and find sensitivity of this search strategy up to heavy quarks masses of $~2.3$ TeV. The analysis is not very sensitive to the mass of the heavy Higgses. Thus, the reach also represents sensitivity of to heavy Higgs masses up to the threshold to decay with a top quark, $~2.1$ TeV.
\subsubsection{$\tau$-jet signatures}
\begin{figure}[t]
\centering
\includegraphics[scale=0.5]{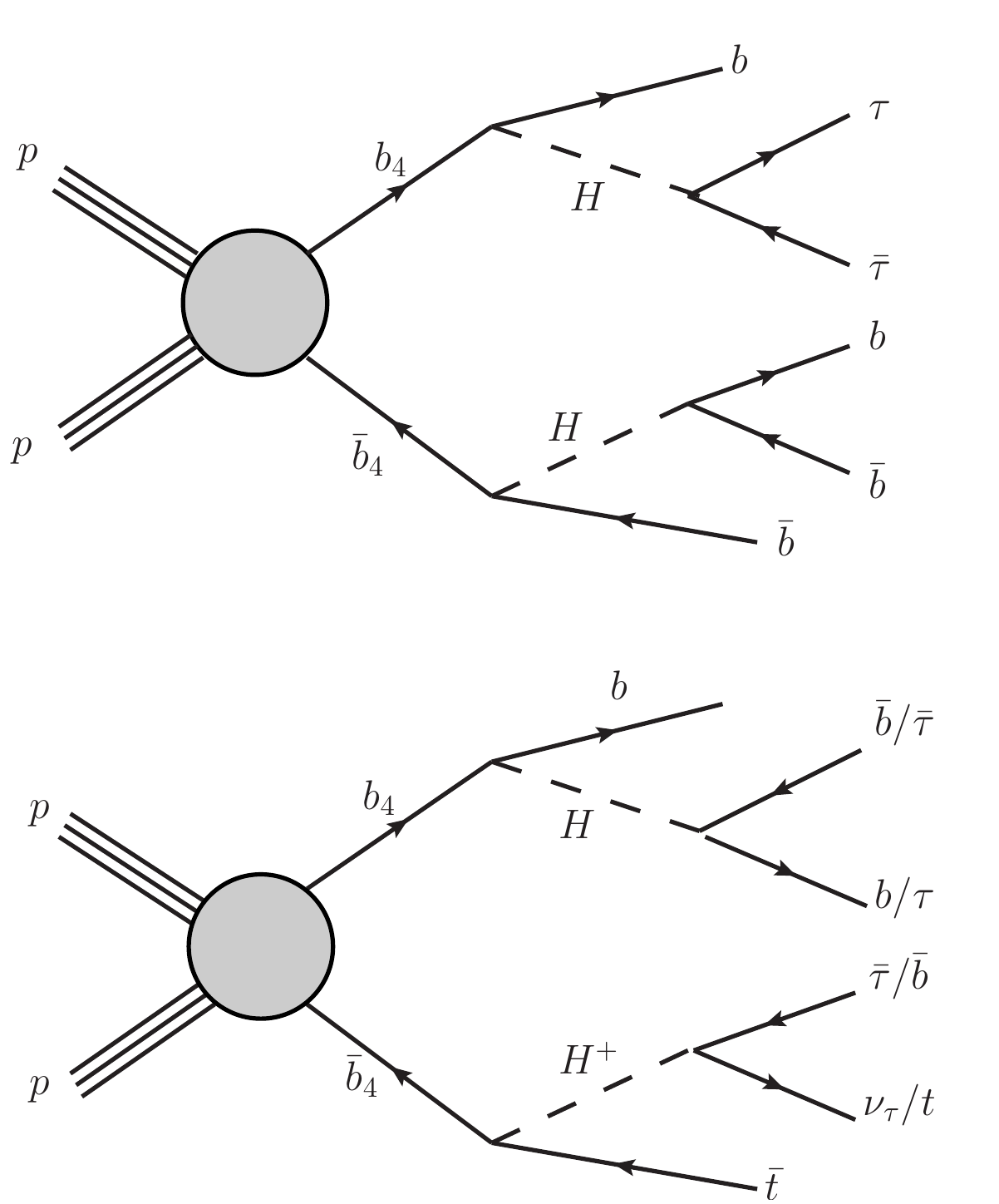}~~~~~~~~~
\includegraphics[scale=0.5]{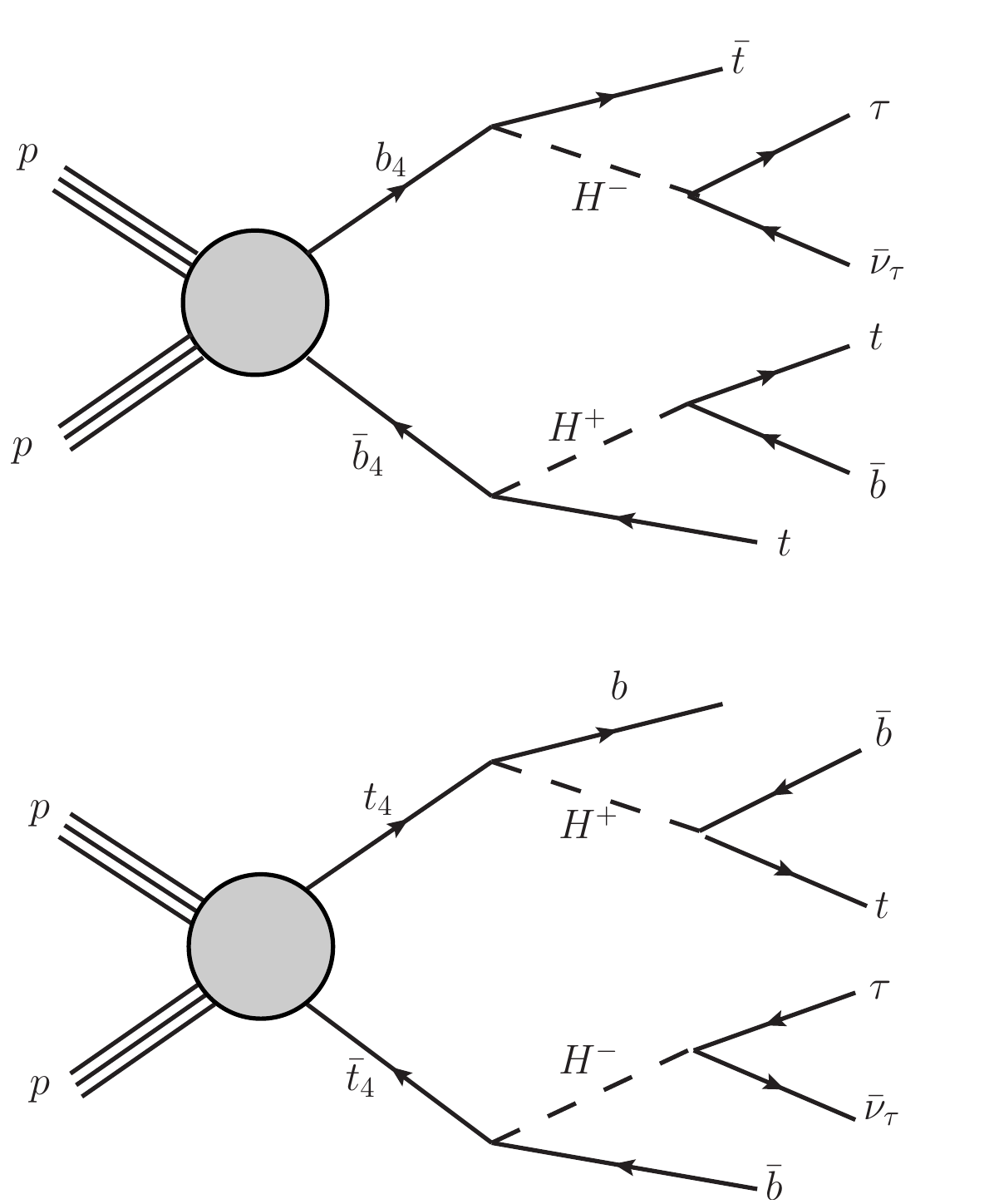}
\caption{Subdominant cascade decays of vectorlike quarks leading to $4b+2\tau$ or $4b+1\tau$ final states in a 2HDM type-II from possible decays of a bottom-like new quark through neutral and charged Higgses or both (left and top right), and the decay of a top-like new quark through charged Higgses (bottom right). This figure is taken from~\cite{Dermisek:2021zjd}.
}
\label{fig:diags}
\end{figure} 
Final states of vectorlike quark cascade decays involving $\tau$ leptons will yield smaller rates in a 2HDM type-II compared the $6b$ signatures discussed in the previous section due to the fact that the typical size of heavy Higgs branching ratios in this case are expected to be ${\rm BR}(H\to \tau\bar{\tau})\simeq{\rm BR}(H^{\pm}\to \tau\nu_{\tau})\simeq 10 \%$. In Fig.~\ref{fig:diags}, we show the resulting decay topologies when one of the neutral Higgses in the decay chain decays to a pair of $\tau$ leptons or a heavy charged Higgs decays to a $\tau$-$\nu_{\tau}$ pair.

Despite the lower production rate of these signatures, this search strategy does not suffer from such huge QCD backgrounds as in the $6b$ analysis. In fact, the situation is not unlike the standard heavy Higgs searches where the best limits come from the $H\to \tau\tau$ channel for the same reasons. In similar fashion as the $6b$ analysis once the decay of the top quark is considered it is clear that all of the resulting signatures can be captured by requiring multiple $b$- and $\tau$-tagged jets in the final state. In these analyses, large hadronic energy and reconstruction of top quark decays present the most useful kinematic variables~\cite{Dermisek:2021zjd}. The latter is constructed by forming the minimum of the invariant mass between a $\tau$-tagged jet and the set of $b$-tagged jets
\begin{equation}
m_{\tau b}=\underset{i}{\text{min}}\sqrt{(p(\tau) + p(b_{i}))^{2}}.
\end{equation}

In Fig.~\ref{fig:BR_bounds}, we show the results of the analyses presented in~\cite{Dermisek:2021zjd} showing the 95\% CL sensitivity of vectorlike quark branching ratios to heavy Higgses using the $\tau$-jet signatures. The solid orange line shows the sensitivity of the $b_{4}\to H b$ branching ratio in this analysis, while the purple and orange dashed lines show the sensitivity to the top-like and bottom-like vectorlike quark decays to the charged Higgs, respectively. The dot-dashed line shows the sensitivity when all decays through neutral and charged Higgses are open. The reach is weaker but comparable to that of the $6b$ analysis. Despite this slight drawback, this channel is complementary to the $6b$ signature offering further information into the dynamics of the heavy Higgs sector and vectorlike quark branching ratios in the event of discovery.

\begin{figure}[t]
\centering
\includegraphics[width=0.6\linewidth]{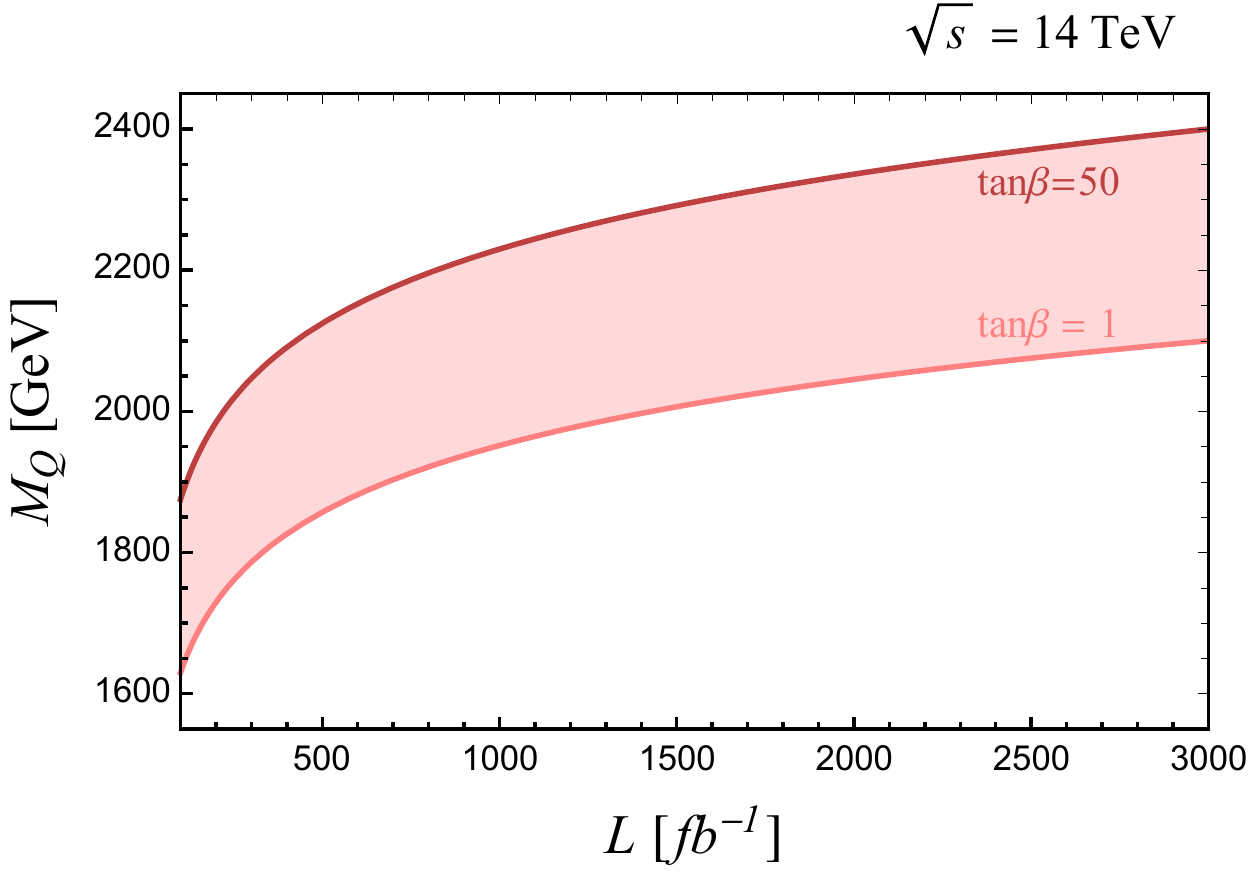}
\caption{Optimal reach of the 14 TeV LHC to vectorlike $SU(2)$ doublet with mass $M_{Q}=m_{t_{4}}\simeq m_{b_{4}}$ decaying to heavy Higgses obtained from the 95\% CL sensitivity on the combined mixed decay of vectorlike quarks with respect to the integrated luminosity. The band shows the variation of the ultimate reach between $\tan\beta = 1$ and $\tan\beta = 50$. This plot is taken from~\cite{Dermisek:2021zjd}.}
\label{fig:MQ_vs_L}
\end{figure}

Finally, we note that in the case that vectorlike top and bottom quarks comprise an $SU(2)$ doublet in the UV theory, their signatures are expected to be produced simultaneously. Thus, the ultimate reach of vectorlike quark decays, either from the $6b$ or $\tau$-jet signatures should be derived from the reach of all the open channels combined. In Fig.~\ref{fig:MQ_vs_L} we show the combined sensitivity to an $SU(2)$ doublet vectorlike quark at the LHC running at 14 TeV as a function of the integrated luminosity. We show the reach from $\tan\beta=1-50$. Thus, by the end of the HL-LHC the expected sensitivity to vectorlike doublet masses is $~2.4$ TeV.

\subsection{Heavy Higgs cascade decays}
\begin{figure}[t]
\centering
\includegraphics[width=0.6\linewidth]{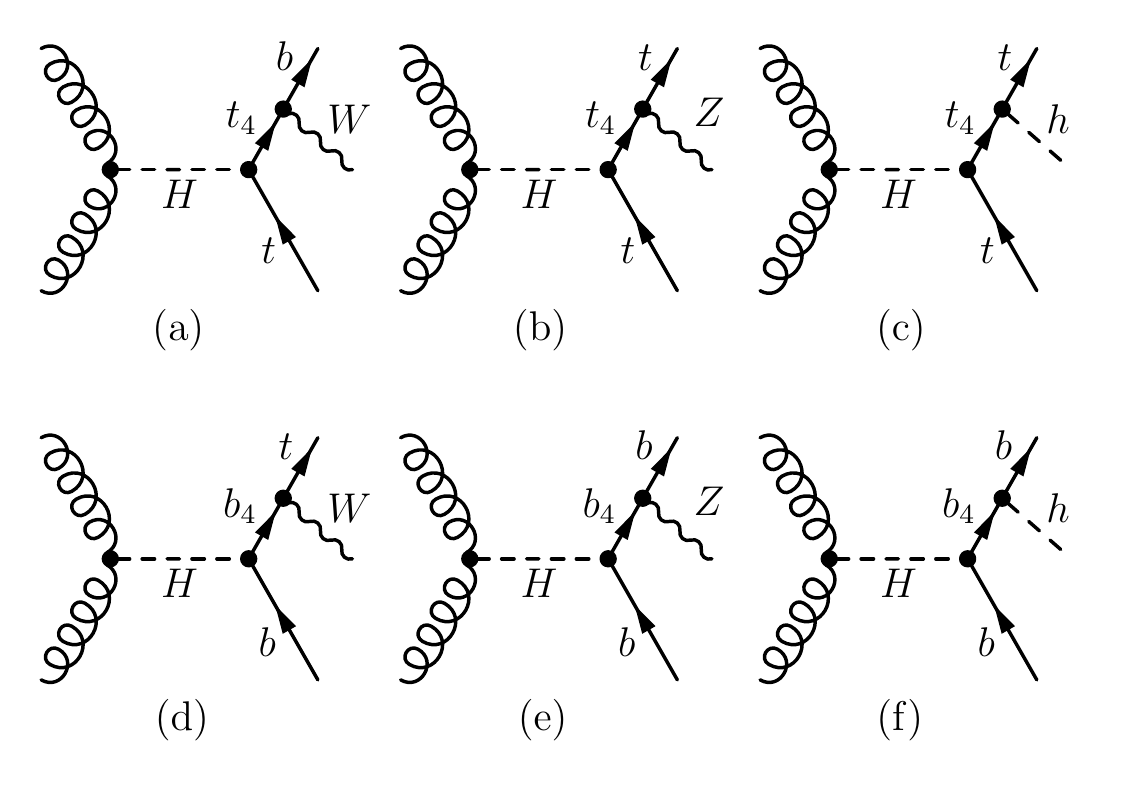}
\caption{\label{fig:topology_Q} 
Possible decay topologies of a heavy neutral Higgs boson through top- and bottom-like vectorlike quarks. Similar decays are obtained with the CP-odd heavy Higgs. This figure is taken from~\cite{Dermisek:2019heo}.
}
\end{figure}
\begin{figure}[t]
\centering
\includegraphics[width=1\linewidth]{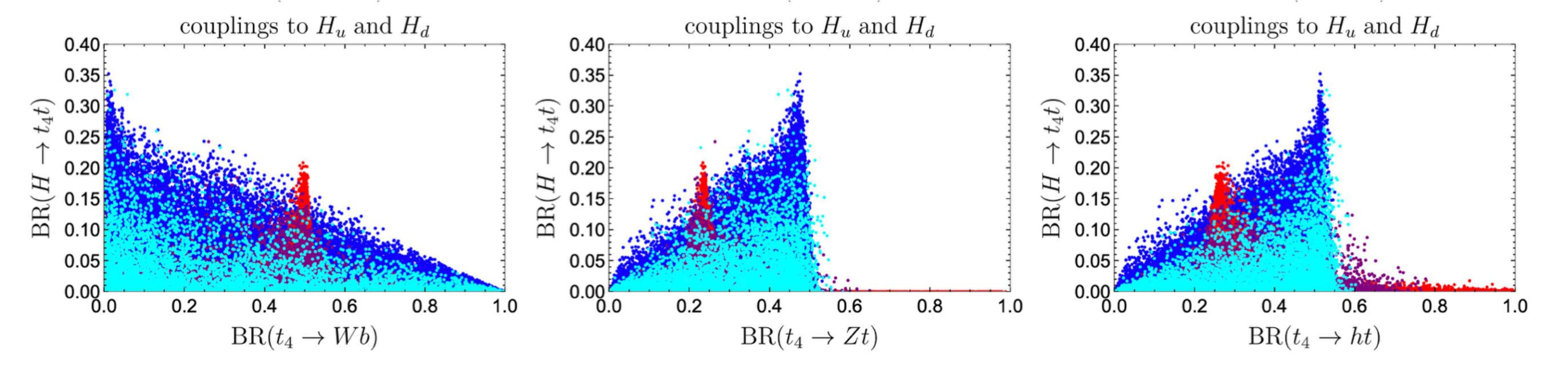}\\
\includegraphics[width=1\linewidth]{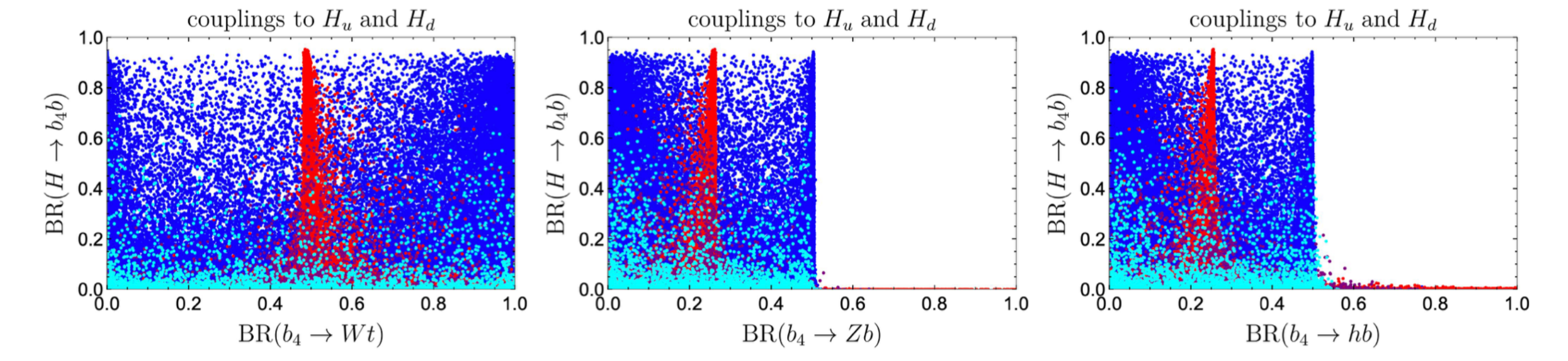}
\caption{\label{fig:HH_cascade_BR} 
Expected branching ratios of vectorlike top- (top) and bottom- (bottom) quarks in a type-II 2HDM to SM bosons compared to the flavor-violating decay of $H$ to a heavy quark and a top or bottom quark. This plot is taken from~\cite{Dermisek:2019heo}.
}
\end{figure}
In the case that heavy Higgses are heavier even than new quarks the possible combined experimental signatures differ in many ways compared to those discussed in the previous section. Signal production involving a heavy neutral Higgs proceeds via the $gg\to H$ cross section which itself varies with respect to both the heavy Higgs mass and $\tan\beta$. Dominant production of vectorlike quarks in the decay chain then follow from the flavor-violating decays of the heavy Higgs to $t_{4}t$ or $b_{4}b$ pairs. In Fig.~\ref{fig:topology_Q}, we show the possible decay chains along these lines including the subsequent decays of vectorlike quarks to SM bosons.

In Fig.~\ref{fig:HH_cascade_BR}, we show the comparison of vectorlike quark branching ratios to SM bosons compared to the flavor-violating decays of the heavy neutral Higgs to $t_{4}t$ (top) and $b_{4}b$ (bottom) pairs. Coupling to both Higgs doublets are varied up to unity. Red, purple, cyan, and blue points denote scenarios where new quarks are $>95\%$ singlet-like, $[50,95]\%$ singlet-like, $[50,95]\%$ doublet-like, and $>95\%$ doublet-like, respectively. The pattern of vectorlike quark branching ratios follows that expected by the Goldstone boson equivalence theorem. For further discussion of both vectorlike quark and heavy Higgs decays in this limit see~\cite{Dermisek:2019vkc,Dermisek:2019heo}.

\begin{figure}[t]
\centering
\includegraphics[width=0.6\linewidth]{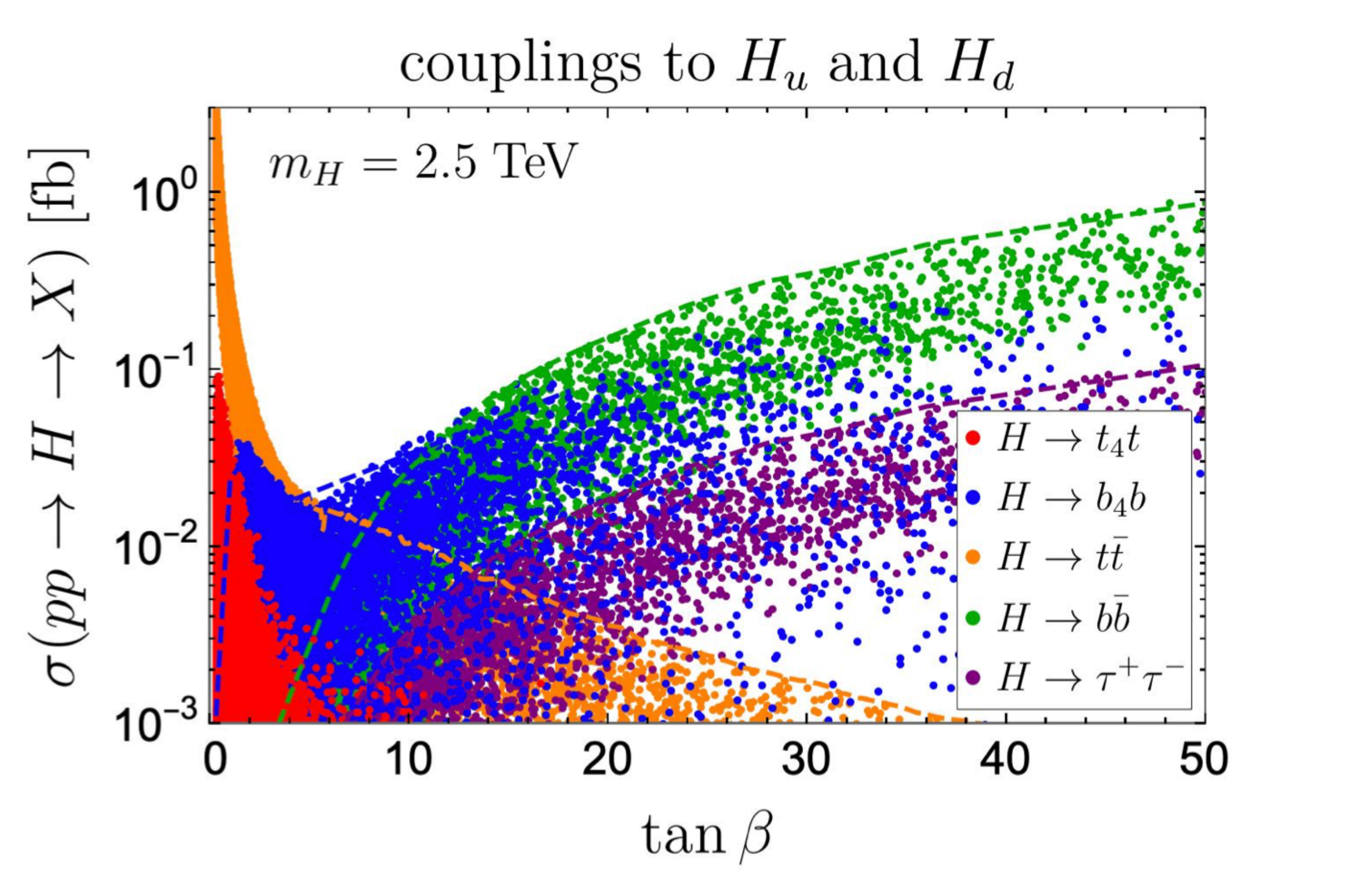}
\caption{\label{fig:cascade_xsection} 
Production cross section involving the flavor-violating decays of a heavy neutral Higgs to $t_{4}t$ or $b_{4}b$ pairs (red and blue) compared to the dominant flavor conserving decays in a type-II 2HDM. This plot is taken from~\cite{Dermisek:2019heo}.
}
\end{figure}

The flavor-violating heavy Higgs decays, $H\to t_{4}t$ and $H\to b_{4}b$, can be at least 10\%, particularly for low to medium $\tan\beta$ and $\tan\beta >0.8$, respectively, and can be as large as 40\% and 95\%, respectively. Combined with the heavy Higgs production, the resulting rates can be possibly orders of magnitude larger than the typical searches for single production of vectorlike quarks. In Fig.~\ref{fig:cascade_xsection}, we show the cross sections for $pp\to H\to t_{4}t/b_{4}b$ (red and blue) compared to those obtained from the dominant flavor-conserving decays of the heavy Higgs for $m_{H}=2.5$ TeV in a type-II 2HDM. Couplings to both Higgs doublets are allowed up to unity. The $pp\to H\to t_{4}t$ signal can be largest at low $\tan\beta$ whereas the $pp\to H\to b_{4}b$ can be large and small or large $\tan\beta$ and can even dominate at medium $\tan\beta$.

Examples of simple cut-based analyses of some of these decay chains was presented in~\cite{Dermisek:2019heo}. It was found that the HL-LHC with 3 ab$^{-1}$ luminosity will have sensitivity to cross sections $pp\to H\to t_{4}t\to Z tt \gtrsim 0.19$ fb and $pp\to H\to b_{4}b\to Z bb \gtrsim 0.16$ fb. We note that while a dedicated analysis is still lacking from the literature, the SM backgrounds are significantly smaller for these channels compared to the typical search channels for single production of vectorlike quarks but also the typical search channels for the flavor conserving decays $H\to bb$ and $H\to tt$.

\section{vectorlike lepton signatures}
\label{sec:VLL}

Vectorlike leptons which mix to SM leptons will likewise have distinct signatures at the LHC.
One of the main production channels of the vectorlike leptons at the LHC is the Drell-Yan process which leaves multi-lepton and missing $E_T$ final states~\cite{Dermisek:2014qca}.
However, unlike the vectorlike quarks produced by QCD processes, the production cross sections of the vecorlike leptons through the electroweak gauge interactions are much smaller than 1pb for the leptons heavier than 300 GeV at the LHC run 3~\cite{CMS:2022nty}. In 2HDM, new processes involving the production of a heavy Higgs boson can avoid this suppression. Hence, we focus on the production of the vectorlike leptons through heavy BSM Higgs cascade decay in this white paper.

\begin{figure}[h]
\centering
\includegraphics[width=0.6\linewidth]{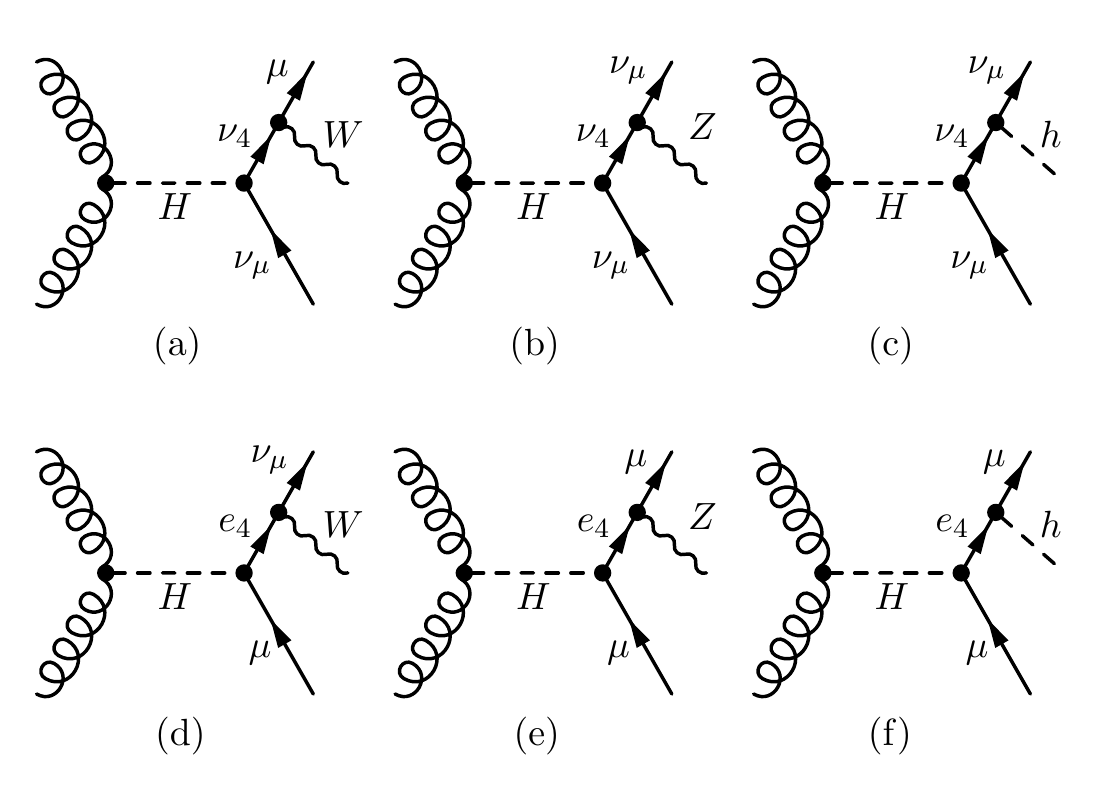}
\caption{
Topologies of a heavy Higgs (either CP even or odd) cascade into vectorlike leptons taken from Ref.~\cite{Dermisek:2015hue}.
\label{fig:topology} 
}
\end{figure}

A prominent feature of BSM models with vectorlike leptons which mix with the SM muon is that the longstanding anomalous magnetic moment of the muon can be explained by the chiral enhancement~\cite{Kannike:2011ng,Dermisek:2013gta} and~\cite{Dermisek:2014cia}.
Interestingly, it is found that additional contributions to the magnetic moment from heavy Higgs bosons in a type-II 2HDM are further enhanced by $\tan^2\beta$ compared to corrections with SM bosons, which allows for very heavy vectorlike leptons and BSM Higgs bosons of masses $\mathcal O (10\,{\rm TeV})$ to explain the anomaly, while satisfying constraints from precision EW data ~\cite{Dermisek:2020cod,Dermisek:2021ajd}.
Motivated by this possibility, we study signals for vectorlike leptons and heavy Higgses at the LHC which may appear when new leptons mix with the muon.
Nevertheless, we believe the same methodology is applicable to other scenarios where vectorlike leptons mix with either the SM electron or tau, as well as other types of Higgs extensions.

\begin{figure}[h]
\centering
\includegraphics[width=0.46\linewidth]{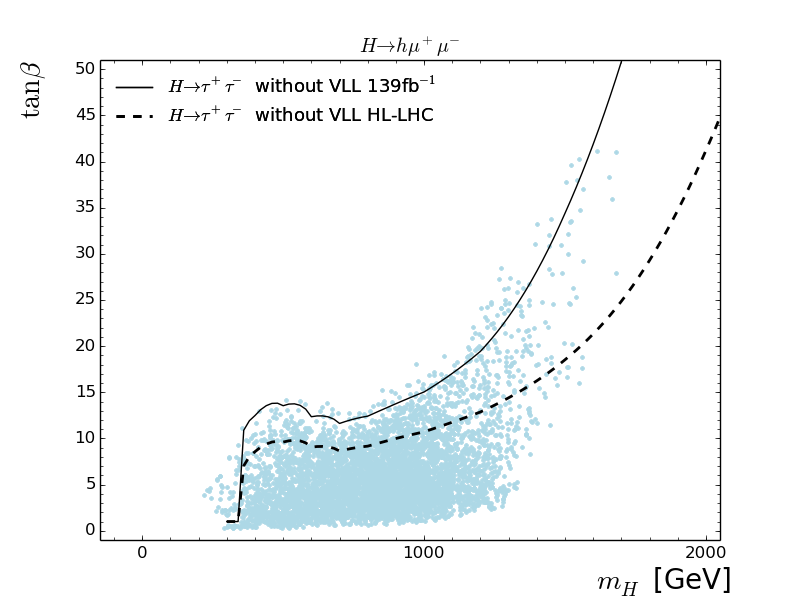}
\includegraphics[width=0.46\linewidth]{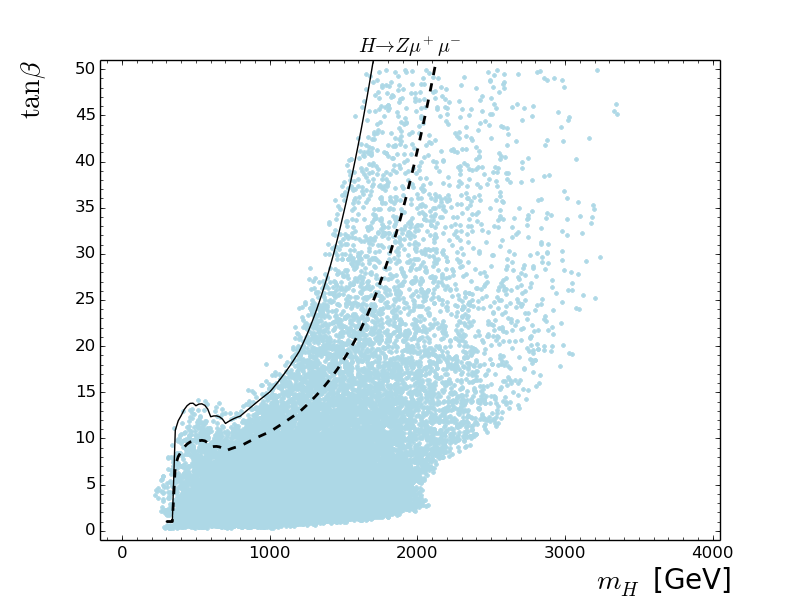}
\caption{
Expected experimental sensitivities (light blue dots) of the cascade decays $H \to e_4^\pm \mu^\mp \to h \mu^+ \mu^- \to b \bar b \mu^+ \mu^-$ (left) and $H \to e_4^\pm \mu^\mp \to Z \mu^+ \mu^-\to b \bar b \mu^+ \mu^-$ (right) in the $m_H - \tan\beta$ plane at the HL-LHC with 3 ab$^{-1}$ integrated luminosity.
The black solid and dashed lines indicate the corresponding reach of the $H \to \tau^+ \tau^-$ search in the absence of vectorlike leptons and in the alignment limit of the 2HDM type-II with the current data (139 fb$^{-1}$) and the HL-LHC full data, respectively.
\label{fig:mHtanbe-2018} 
}
\end{figure}

The main cascade processes we consider are depicted in Fig.~\ref{fig:topology} where $e_4$ ($\nu_4$) refers to the lightest charged (neutral) heavy lepton mass eigenstates. Details of the model can be found in Ref.~\cite{Dermisek:2015vra,Dermisek:2015oja}.
In a large range of parameter space, branching ratios of heavy Higgses to vectorlike leptons can be sizable or even dominant while satisfying the constraints from EW precision observables, heavy Higgs boson searches, and multilepton events stated above.
Depending on the decay products of the EW gauge bosons or the SM Higgs boson, multi-resonance signals can be possibly observed at the LHC, which increases the detectability of the signals in Fig.~\ref{fig:topology}.
For example, one of the cleanest signatures is $H \to e_4^\pm \mu^\mp \to h \mu^+ \mu^-$ with $h \to \gamma \gamma$ or $b \bar b$, which shows sensitivity even up to the branching ratio $\mathcal O(1\%)$ when $m_H$ is below the $t \bar t$ threshold~\cite{Dermisek:2016via}.

It is expected that searches for heavy Higgs cascade decays via vectorlike leptons will also increase the experimental sensitivities of the Higgs parameters such as heavy Higgs mass and $\tan\beta$.
We show the estimated experimental sensitivities (95\% CLs) at the HL-LHC with 3 ab$^{-1}$ integrated luminosity data in Fig.~\ref{fig:mHtanbe-2018} with light blue scattered dots by considering the heavy CP even Higgs cascade processes $H \to e_4^\pm \mu^\mp \to h \mu^+ \mu^-$ (left) and $H \to e_4^\pm \mu^\mp \to Z \mu^+ \mu^-$ (right) where $h$ and $Z$ bosons decay to $b \bar b$ with sizable branching ratios allowing the new Yukawa couplings between vectorlike leptons and Higgs bosons up to 1. (Here, we assume the other heavy Higgs bosons are much heavier.)
In obtaining the results, we kept the selection cuts in Ref.~\cite{Dermisek:2016via} including an {\it off-$Z$} cut ($|m_{\mu^+ \mu^-} - M_Z| > 15$ GeV) to reject a large portion of the background events from $Z \to \mu^+ \mu^-$. Background events are estimated by a simple rescaling assuming that the rejection efficiencies are the same.
For the signal, we take a flat signal acceptance of 30\% for $m_H \le 800$ GeV, motivated from the values in Ref.~\cite{Dermisek:2016via}, and 50\% for $m_H > 800$ GeV since harder di-muon productions will help the signal pass the {\it off-$Z$} cut more efficiently.
Also, we take a flat detector efficiency, 50\%, for the whole parameter range.
The light blue scattered dots in Fig.~\ref{fig:mHtanbe-2018} satisfy various constraints from electroweak precision measurements~\cite{Zyla:2020zbs}, multi-lepton with missing $E_T$ signals from Drell-Yan pair production of vectorlike leptons~\cite{Dermisek:2014qca}, heavy Higgs searches in the $H \to \gamma \gamma$~\cite{CMS:2016kgr,ATLAS:2017ayi} and the $H \to \tau^+ \tau^-$ channels~\cite{ATLAS:2020zms}.
We see that the HL-LHC can reach Higgs masses up to $m_H \sim 1.7$ TeV for the $H \to e_4^\pm \mu^\mp \to h \mu^+ \mu^-$ process and $m_H \sim 3.3$ TeV for $H \to e_4^\pm \mu^\mp \to Z \mu^+ \mu^-$.

\begin{figure}[t]
\centering
\includegraphics[width=0.38\linewidth]{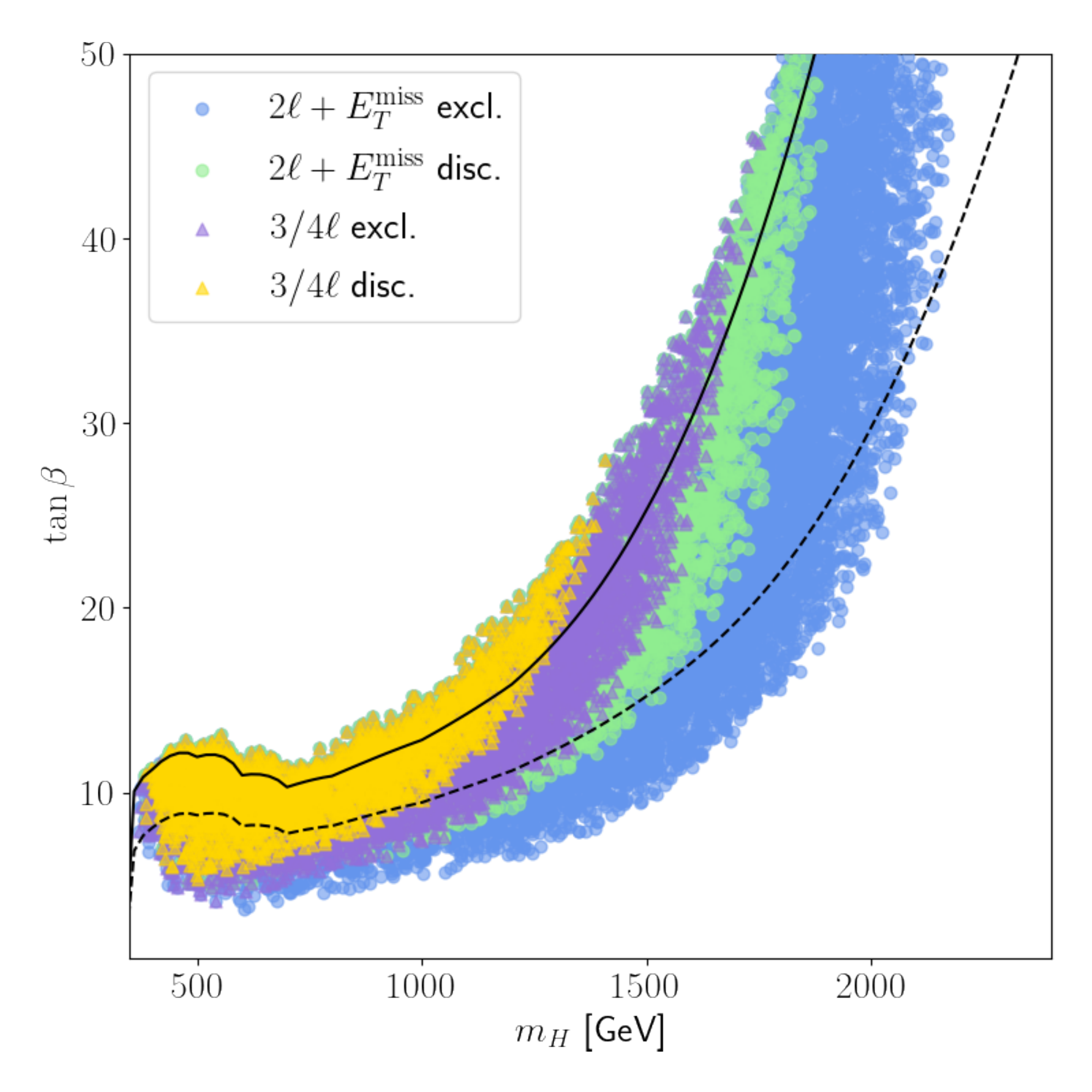}
\includegraphics[width=0.38\linewidth]{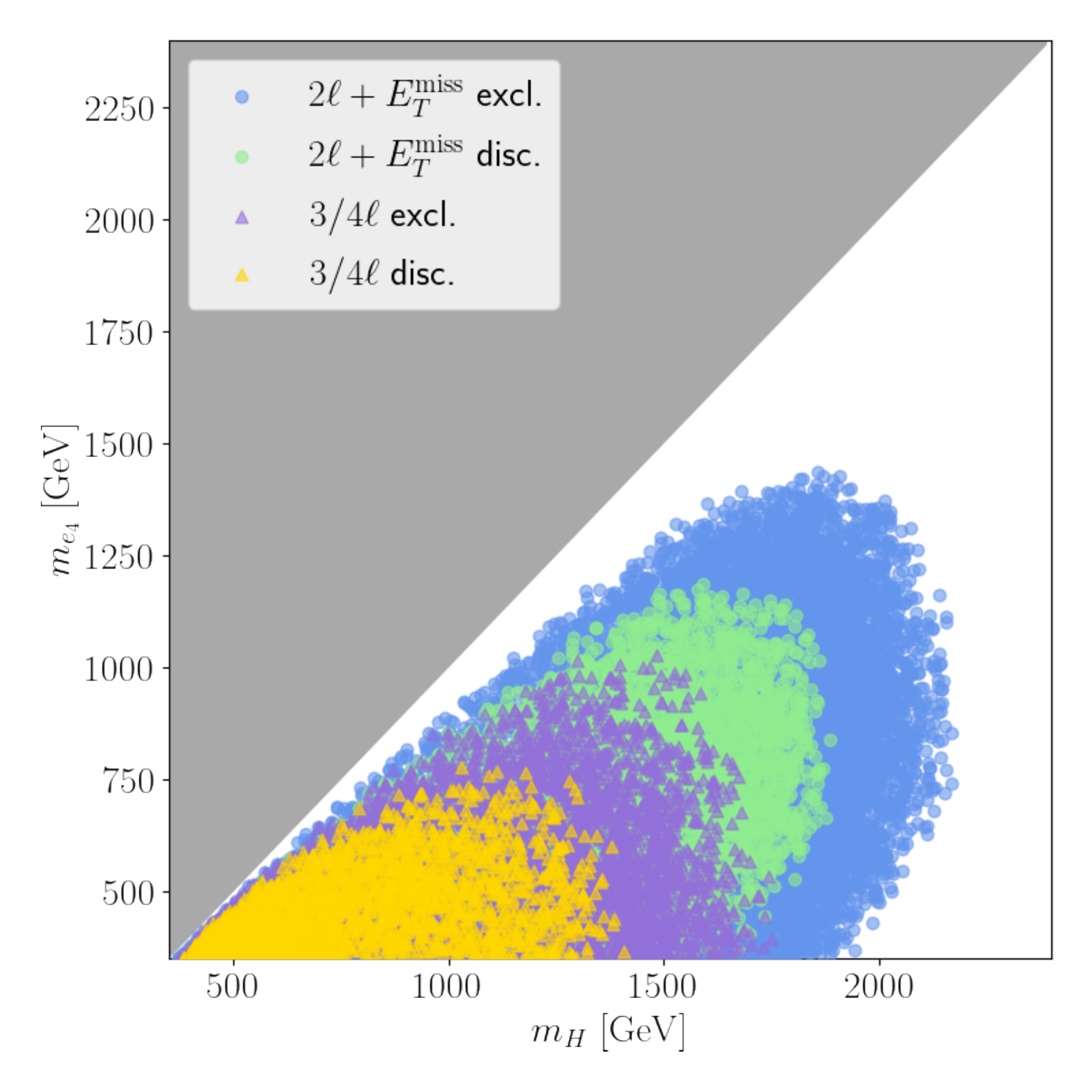}
\caption{
Discovery and exclusion reach of the HL-LHC for multi-lepton signatures of heavy Higgs cascade decays.
The yellow and purple dots correspond to the expected parameter region which we expect can be discovered and excluded, respectively, by the 3/4 lepton channel search as in Ref.~\cite{ATLAS:2021wob}, while the green and blue dots show the respective ranges using the 2 leptons plus missing transverse energy channel search as in Ref.~\cite{ATLAS:2019lff}.
The points satisfy the current constraints from the EW precision measurements, and the $H/A \to \tau^+ \tau^-$ search at the LHC. 
In the left panel, the black solid (dashed) line is the exclusion limit by the $H/A \to \tau^+ \tau^-$  search with the current (projected HL-LHC) data in the absence of the vectorlike leptons and in the alignment limit of the two Higgs doublet model type-II.
In the right panel, we show the expected experimental reach in $m_H - m_{e_4}$ plane at the HL-LHC.
\label{fig:mHlepton} 
}
\end{figure}

In order to show the advantages of cascade decay searches with heavy leptons over conventional heavy Higgs searches, we also show the experimental reach of $H \to \tau^+ \tau^-$ in the absence vectorlike leptons and in the alignment limit of the 2HDM type-II.
The black solid line shows reach from current data with 139 fb$^{-1}$ ($H \to \tau^+ \tau^-$ without VLL 139 fb$^{-1}$) and the dashed line shows the corresponding result for the HL-LHC with 3 ab$^{-1}$ integrated luminosity ($H \to \tau^+ \tau^-$ without VLL HL-LHC). We see that the expected sensitivities of heavy Higgs cascade decays are able to probe lower values of $\tan\beta$ for a given $m_{H}$ compared to the conventional $H \to \tau^+ \tau^-$ searches. This feature persists for heavy Higgs masses $m_{H}\simeq 1.5$ TeV in the $H \to h \mu^+ \mu^-$ channel, and even up to $m_{H}\simeq 3.5$ TeV for $H \to Z \mu^+ \mu^-$.

In addition to the $h\to \bar{b}b$ and $Z\to \bar{b}b$ channels associated with the $\mu^+ \mu^-$ in the cascade process, we consider the leptonic decay channels with cleaner particle identifications.
The sensitivities of these multi-lepton final states can be recasted using various existing searches for BSM particles in supersymmetric theories or seesaw models, for example the $3/4\ell$ analysis of ~\cite{ATLAS:2021wob} and the $2\ell + E_T^{\rm miss}$ of~\cite{ATLAS:2019lff}.
Figure~\ref{fig:mHlepton} shows the expected reaches at the HL-LHC 
for a singlet-like vectorlike lepton produced in the cascade decay mode, allowing the new Yukawa couplings between the vectorlike leptons and the Higgses up to 1.
The yellow (purple) dots correspond to the expected range of heavy Higgs and new lepton masses which can be discovered (excluded) in the $3/4\ell$ channel, 
while the green (blue) dots show the corresponding range of masses in the $2\ell + E_T^{\rm miss}$ channel. 
Here, we include the production and decay of both the CP-even and CP-odd neutral Higgs bosons.  
As in the hadronic channel searches, the points satisfy the current constraints from the EW precision measurements 
and the $H/A \to \tau^+ \tau^-$ search at the LHC. 
The leptonic channels provide a complementary approach in probing 
both extended Higgs and matter sectors.
Moreover, the analysis results can be readily applicable in other BSM models, 
such as a new gauge boson which decays to a vectorlike lepton and a SM lepton.  
A more detailed analysis will be presented in future work~\cite{future}.

\section{Discussion and summary}
Combined signatures of multiple new particles are expected in models of new physics which provide a UV completion of the SM. Typical search strategies focus on the production of individual particles and their subsequent decay to SM particles. However, the final states obtained with this strategy in mind may be suppressed in certain regions of parameters of the UV model due to small branching ratios and the combined signatures become the dominant production mode of new physics at the LHC. In the event of discovery, a full profile of possible signatures would be of value to disentangle the underlying dynamics of physics above the TeV scale. In particular, we motivate searches for combined signatures in models with extra Higgs bosons and vectorlike fermions.

We have discussed promising final states resulting from the cascade decays of vectorlike quarks to charged and neutral Higgs bosons in a type-II 2HDM. Apart from the expectation of higher rates over the decays of vectorlike quarks to $W$, $Z$, or $h$, there are kinematic advantages of cascade decays that allow for simple cut-based analyses. We introduced search strategies with this in mind based on tagging of multiple, widely separated $b$-jets and large hadronic energies. Dominant channels are found to be those where heavy Higgses decay to top and bottom quarks. However, including the possibility of heavy Higgses decaying to $\tau$-leptons results in a similar reach while having the advantage of simpler background modeling. Combining these approaches we found that the HL-LHC can probe vectorlike quark masses $2-2.4$ TeV. Only slightly lower reach of heavy Higgs masses is expected as the sensitivity of these searches does not depend on the mass of charged or neutral Higgses, as long as the decay chain is kinematically open.

In the case that heavy Higgses are heavier than new quarks, we have presented expected rates of cascade decays of a heavy neutral Higgs to single production of new top- and bottom-like new quarks. A dedicated analysis of the reach of new quark and Higgs masses at the LHC is still lacking for these channels. However, it is expected that the relevant backgrounds for the resulting final states are significantly smaller than those compared to the typical searches for single production of vectorlike quarks. In contrast to the searches based on heavy quark cascade decays, it should be mentioned that the Higgs cascade decay channels are more model dependent in the sense that the expected size of the production rate is sensitive to the size of the coupling in the model which mixes heavy quarks to SM quarks.

Searches for heavy Higgs cascade decays with heavy leptons also present promising search strategies. Motivated by the possibility to probe solutions to the measured discrepancy of the muon anomalous magnetic moment we have proposed several cascade decay channels resulting from heavy Higgs production and vectorlike leptons mixing with the muon. Simple estimates based on the analysis of Ref.~\cite{Dermisek:2016via}, show that the possible channels $H \to e_4^\pm \mu^\mp \to (Z\to\bar{b}b)\mu^+ \mu^-$ and $H \to e_4^\pm \mu^\mp \to (h\to\bar{b}b)\mu^+ \mu^-$ will have sensitivity at the HL-LHC beyond what is currently expected from the conventional $H\to \tau^{+}\tau^{-}$ searches with respect to the reach in $m_{H}$ vs $\tan\beta$. Further, multi-lepton final states offer a similar reach in $m_{H}$ vs $\tan\beta$ and can be easily recasted using existed searches.

It should be noted that while we have explored various search strategies for cascade decays involving heavy Higgs bosons and vectorlike quarks and leptons in a type-II 2HDM, similar channels would appear in other models. For instance, the channels we present for charged and neutral Higgs bosons would be identical to those involving new $W^\prime$ and $Z^\prime$ bosons (assuming the relevant couplings in the model exist). Thus, up to rescaling of the appropriate branching ratios, our results can also utilized to estimate the expected sensitivities for a broad classes of searches for new physics at the LHC.
\bibliography{ref}

\end{document}